\documentstyle[aps,preprint,epsfig,axodraw,amsmath,amssymb,graphicx,mathrsfs]{revtex}
\newcommand{\BRbsgamma}{\textrm{BR}(b \to s \gamma)}
\newcommand{\bsgamma}{b \to s \gamma}
\newcommand{\GeV}{\textrm{GeV}}
\newcommand{\Neuone}{ \tilde{\chi}^0_1 }
\newcommand{\Neutwo}{ \tilde{\chi}^0_2 }
\newcommand{\Chaone}{\tilde{\chi}^{\pm}_1}

\begin{document}
\preprint{TU-610}
\title{Higgs mass and $\bsgamma$ constraints on SUSY models 
with no-scale boundary condition}

\author{Shinji Komine\footnote{komine@tuhep.phys.tohoku.ac.jp}}
\address{Department of Physics, Tohoku University,
Sendai 980-8578, Japan}
\date{\today}

\maketitle
\begin{abstract}

No-scale structure of the K\"ahler potential is obtained 
in many types of supersymmetric models.
In this paper, phenomenological aspects of these models are investigated
with special attention to the current Higgs mass bound at LEP and 
$b \to s \gamma$ result at the CLEO. 
When the boundary condition is given at the GUT scale and gaugino masses are 
universal at this scale, very narrow parameter region is allowed only for 
positive Higgsino mass region if R-parity is conserved. 
The negative Higgsino mass case is entirely excluded.
On the other hand, relatively large parameter region is allowed 
when the boundary condition is given above the GUT scale, 
and Tevatron can discover SUSY signals for the positive Higgsino mass case.
The no-scale models with Wino, Higgsino or sneutrino LSP are also considered.
We show that the Higgs mass constraint is important for the Higgsino LSP case,
which requires the LSP mass to be larger than about 245 GeV.

\end{abstract}
\clearpage

\section{Introduction}
Supersymmetry (SUSY) is one of the most attractive extensions of 
the standard model.
This symmetry solves the naturalness problem and predicts gauge coupling 
unification at the GUT scale $M_{GUT} \simeq 2 \times 10^{16} \GeV$.
It also predicts the existence of superpartner of 
the standard model (SM) particles.
From the naturalness argument, their masses should be below TeV range,
hence these particles will be discovered at Tevatron or 
Large Hadron Collider (LHC).

Mechanisms of SUSY breaking and its mediation to the minimal supersymmetric 
standard model (MSSM) sector are one of the most important problems
in the SUSY phenomenology.
In many models, this dynamics is related to high energy physics 
far above the electroweak(EW) scale, e.g., GUT scale or Planck scale. 
Once the mechanism is specified, mass spectrum and flavor structure of 
SUSY particle at the EW scale can be determined 
by a small number of parameters.
Hence it may be possible to confirm or exclude the mechanism by direct search 
or flavor-changing-neutral-current (FCNC) experiments in near future.

If SUSY breaking is mediated by gravity, 
the structure of SUSY breaking masses of scalars are determined by 
K\"ahler potential. 
In the present paper, we focus on the no-scale type K\"ahler potential, 
in which the hidden sector and the observable sector are separated as follows:
\begin{eqnarray}
  e^{-K/3} = f_{hid}(z,z^*) + f_{obs}(\phi, \phi^*),
  \label{eq:NoScaleKahler}
\end{eqnarray}
where $z$ and $\phi$ are hidden sector fields and 
observable sector fields, respectively.
This type of K\"ahler potential is originally investigated in Ref.
\cite{OriginalNoScale} with $f_{hid}(z,z^*) = z + z^*$ and 
$f_{obs}(\phi, \phi^*) = \sum_i \phi_i \phi_i^*$.
Characteristic features of the K\"ahler potential eq.(\ref{eq:NoScaleKahler}) 
is that 
all scalar masses and trilinear scalar couplings (A-terms) 
vanish as the cosmological constant vanishes\cite{IKYY}.
The only source of SUSY breaking is gaugino masses.
Hence this scenario is highly predictive, 
various phenomenological consequences are obtained with a few parameters.
The separation in eq.(\ref{eq:NoScaleKahler}) implies that couplings of the 
hidden sector and the observable sector is flavor blind, 
and contributions of SUSY particles to FCNC are suppressed.
Therefore this K\"ahler potential is also interesting 
from the viewpoint of the SUSY flavor problem. 

The no-scale structure of the K\"ahler potential is obtained in various models.
It has been shown that in some classes of string theory, 
for example weakly coupled $E_8 \times E_8$ heterotic string theory, 
K\"ahler potential becomes the no-scale type\cite{StringModel,Mtheory}.
If the hidden sector and the observable sector are separated in the superspace 
density in the supergravity Lagrangian, the K\"ahler potential is indeed 
given as in eq. (\ref{eq:NoScaleKahler}).
In the two cases, the gaugino masses can be induced 
if the hidden sector fields couple to the gauge multiplets 
via the gauge kinetic function.
Recently it has been pointed out that the form eq.(\ref{eq:NoScaleKahler}) is 
realized naturally in a five-dimensional setting with two branes, namely, 
sequestered sector scenario\cite{sequesterd}.
In this scenario, the hidden sector fields live on one brane and 
the visible sector fields live on the other.
It has been shown that the  form of the K\"ahler potential of the effective 
theory obtained by dimensional reduction is indeed 
eq.(\ref{eq:NoScaleKahler})\cite{LutySundrum}.
If the SM gauge fields dwell in the bulk, 
gaugino mediate the SUSY breaking on the hidden sector brane to the visible 
sector brane and the no-scale boundary condition is given 
at the compactification scale of the fifth dimension 
(gaugino mediation \cite{GauginoMediation,SchmaltzSkiba}).

In the no-scale scenario, degrees of freedom of SUSY particle mass spectrum 
is limited because only non-zero soft SUSY breaking masses are 
gaugino masses and Higgs mixing mass $B$ at the energy scale 
where the boundary condition is given.
Hence phenomenological aspects of this scenario have been investigated 
in the literature, mainly focusing on the mass spectrum.
Direct search bounds and the cosmological constraint 
(i.e., a charged particle can not be the LSP if the R-parity is conserved) 
were considered and 
allowed region in the parameter space was identified.
For the boundary condition, the following three cases were considered.
First, universal gaugino masses are given at the GUT scale.
In this case, the cosmological constraint is severe and only the region 
$200 \GeV \lesssim M_{1/2} \lesssim 250 \GeV$ and $\tan \beta \lesssim 8$ is 
allowed since stau tends to be light\cite{IKYY,KomineYamaguchi}.
The second case is that universal gaugino masses are given above the GUT scale.
And the third case is that non-universal gaugino masses are given at the 
GUT scale.
In this case Wino, Higgsino or sneutrino can be the LSP.
In the latter two cases, 
it is shown that the cosmological constraint is not severer 
than the first case.  

In the present paper, current limits from the lightest Higgs mass $m_h$ 
and the branching ratio for $b \to s \gamma$ are also used to constrain
the no-scale scenario.
Combining these constraints, we will show that 
almost all the parameter region is excluded when universal gaugino masses are 
given at the GUT scale.
However, when the boundary condition is given above the GUT scale, 
relatively large parameter region is allowed.
We also consider the case that the non-universal gaugino masses are given 
at the GUT scale.
We will show that these constraints are important when the 
Higgsino-like neutralino is the LSP.

This paper is organized as follows.
In section \ref{sec:NoScalseBC}, we review some phenomenological aspects of 
the no-scale models, especially indications of the direct search bounds and 
the cosmological bound. 
In section \ref{sec:HiggsBsgamma}, we further constrain these models from the 
Higgs mass bound and $\BRbsgamma$ result. 
Indications of these bounds for the Tevatron are also discussed. 
Our conclusions are given in section \ref{sec:conclusions}. 

\section{Models with No-scale boundary condition}
\label{sec:NoScalseBC}
In this section, we briefly review phenomenological aspects of SUSY models 
with no-scale boundary condition, mainly indications of the cosmological bound
and direct search limit at LEP 2.   
We consider the following three cases.
\begin{itemize}
\item Universal gaugino masses are given at the GUT scale. Hereafter we call 
this case the minimal scenario.
\item Universal gaugino masses are given above the GUT scale $M_* > M_{GUT}$.
Throughout this paper, we take the minimal SU(5) to be the theory 
above the GUT scale as a typical example.
\item Non-universal gaugino masses are given at the GUT scale.
\end{itemize}
Once one of the above boundary conditions is given, 
mass spectrum of SUSY particles at the EW scale and their contributions to 
FCNC can be calculated.
In this paper we solve the one-loop level RGEs to obtain the soft SUSY 
breaking mass parameters at the EW scale. 
The Higgsino mass parameter $\mu$ is determined by 
potential minimum condition at the one-loop level.

\begin{figure}[tbp]  
  \begin{center}
    \includegraphics[height=7cm,width=9cm,keepaspectratio,clip]
    {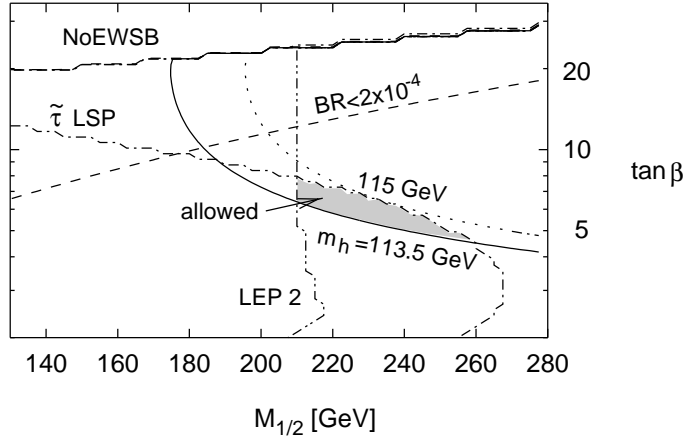}
  \end{center}
  \caption{Constraint in the $M_{1/2} - \tan \beta$ plane 
    for the minimal case with $\mu > 0$.
    In the region above the dash-dotted line, the stau is the LSP.
    The left side of the dash-dot-dotted line is excluded by the upper bound 
    on smuon pair production cross section at LEP. 
    The current Higgs mass bound excludes 
    the region below the solid line. In the region above the dashed line, 
    $\BRbsgamma$ is smaller than the lower limit obtained by the CLEO. 
    The shaded region is allowed. The $m_h = 115 \GeV$ curve is also shown 
    as the dotted line. 
    'NoEWSB' means that radiative breaking does not occur.}
\label{fig:Limitmimposmu}
\end{figure}

\begin{figure}[tbp]  
  \begin{center}
    \includegraphics[height=7cm,width=9cm,keepaspectratio,clip]
    {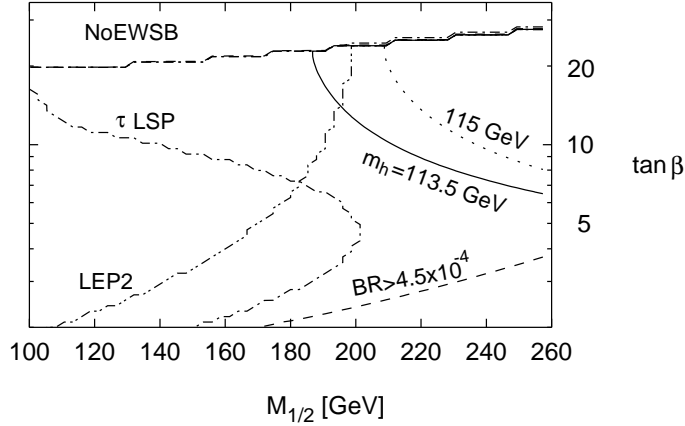}
  \end{center}
  \caption{The same as Fig.\ref{fig:Limitmimposmu} but $\mu < 0$. 
    The region above 
    the dashed line is excluded since $\BRbsgamma$ is larger than the upper 
    limit obtained by the CLEO. The other lines are the same as in 
    Fig.\ref{fig:Limitmimposmu}. Notice that all region is excluded.}
\label{fig:Limitmimnegmu}
\end{figure}

First, we discuss the minimal scenario. In this case, 
the following boundary condition is given at the GUT scale, 
\begin{eqnarray}
  & & m_0^2=0, \quad A_0 = 0,  \nonumber \\
  & &M_1(M_{GUT}) = M_2(M_{GUT}) = M_3(M_{GUT}) = M_{1/2} ~,
\end{eqnarray}
where $m_0$ is the common scalar mass and $A_0$ is universal trilinear 
scalar coupling.
With this boundary condition, 
Bino and right-handed sleptons are lighter than other SUSY particles.
Their masses are approximately,
\begin{eqnarray}
  M_1^2 \simeq 0.18 M_{1/2}^2 , \quad 
  m_{\tilde{e}_R}^2 \simeq 0.15 M_{1/2}^2 -0.23 m_Z^2 \cos 2\beta .
  \label{eq:minimalmNeumtau}
\end{eqnarray}
From eq.(\ref{eq:minimalmNeumtau}) we see that the charged right-handed 
slepton is the LSP if the D-term $0.23 m_Z^2 \cos 2\beta$ is negligible, 
i.e., $M_{1/2} \gtrsim 250 \GeV$.
Hence this parameter region is excluded by the cosmological consideration.
On the other hand, LEP 2 experiments yields the upper bound on the cross 
section for smuon pair production, 
$\sigma ( e^+ e^- \to \tilde{\mu}^+_R \tilde{\mu}^-_R ) < 0.05 \textrm{pb}$ 
for $m_{\tilde{\mu}_R} \leq 98 ~\textrm{GeV}$ and $m_{\tilde{\chi^0_1}}
\leq 0.98 m_{\tilde{\mu}_R} - 4.1 ~\textrm{GeV}$\cite{SUSY2000LEP}, 
so the parameter region 
$ M_{1/2} \lesssim 200 \GeV$ is excluded
In Fig. \ref{fig:Limitmimposmu} and \ref{fig:Limitmimnegmu}, 
allowed region of the parameter space are shown in the 
$M_{1/2} - \tan \beta$ plane.
The regions above the dash-dotted line and the left side of the 
dash-dot-dotted line are excluded by cosmological bound and 
LEP 2 bound on smuon pair production, respectively.
Therefore the minimal scenario is constrained severely.

\begin{figure}[tbp]  
  \begin{center}
    \includegraphics[height=7cm,width=9cm,keepaspectratio,clip]
    {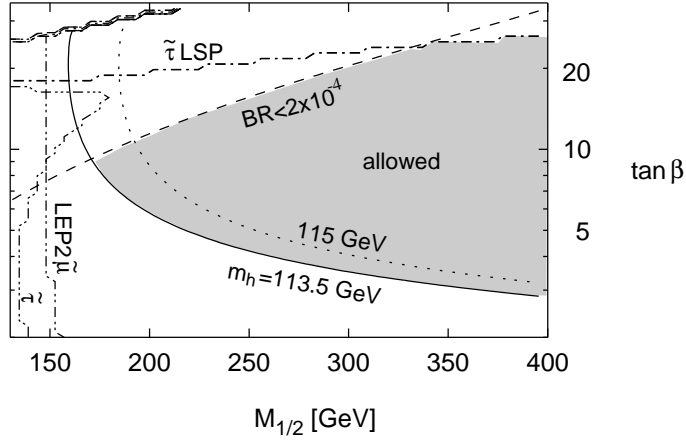}
  \end{center}
  \caption{Constraint in the $M_{1/2} - \tan \beta$ plane 
    for universal gaugino masses, $M_* = 10^{17} \GeV$ 
    and $\mu > 0$. The left sides of the dash-dot-dotted and 
    dash-dot-dot-dotted line are excluded by the upper bound on the smuon and 
    stau pair production cross section, respectively. The other lines are the 
    same as in Fig.\ref{fig:Limitmimposmu} }
\label{fig:LimitMBC1e17posmu}
\end{figure}

\begin{figure}[tbp]  
  \begin{center}
    \includegraphics[height=7cm,width=9cm,keepaspectratio,clip]
    {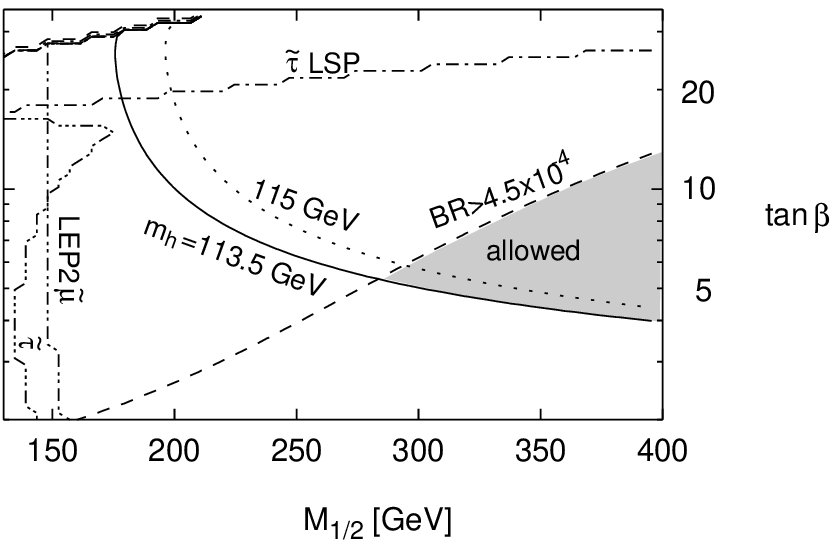}
  \end{center}
  \caption{The same as Fig.\ref{fig:LimitMBC1e17posmu} but for $\mu < 0$.}
\label{fig:LimitMBC1e17negmu}
\end{figure}

Next we see the case that the universal gaugino masses are given 
above the GUT scale.
In the minimal SU(5) case, the right-handed slepton belongs to 10-plet,
so the large group factor makes slepton masses heavier.
For example, when $M_* = 10^{17} \GeV$ the Bino mass and 
the right-handed slepton mass at the weak scale are approximately given,
\begin{eqnarray}
  M_1^2 \simeq 0.18 M_{1/2}^2, \quad 
  m_{\tilde{e}_R}^2 \simeq 0.30 M_{1/2}^2 -0.23 m_Z^2 \cos 2\beta .
\end{eqnarray}
Hence the cosmological constraint is not severe because the stau mass is 
large enough and neutralino is the LSP in the large parameter region
\cite{KMY,PolonskyPomarol,SchmaltzSkiba}.
In the Fig.\ref{fig:LimitMBC1e17posmu} and \ref{fig:LimitMBC1e17negmu}, 
the same figures as in the Fig.\ref{fig:Limitmimposmu} and 
\ref{fig:Limitmimnegmu} are shown.
Unlike in the minimal case, the stau search bound at LEP \cite{SUSY2000LEP}
is also plotted
because mass difference between $\Neuone$ and $\tilde{\tau}$ is larger than 
in the minimal case and it can be stronger than the smuon search.
From these figures we see that the $\tilde{\tau}$ LSP is avoided 
unless $\tan \beta$ is larger than about 20.

\begin{figure}[tbp]  
  \begin{center}
    \includegraphics[height=7cm,width=9cm,keepaspectratio,clip]
    {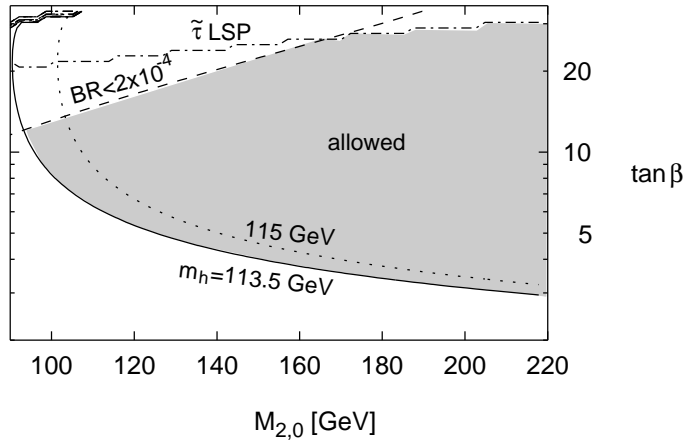}
  \end{center}
  \caption{Constraint in the $M_{2,0} - \tan \beta$ plane for 
    the Wino LSP case, $M_{1,0}/M_{2,0}=4$, $M_{3,0}/M_{2,0}=2$, 
    $M_* = M_{\textrm{GUT}}$ and $\mu > 0$.}
\label{fig:Limit12R432R2posmu}
\end{figure}

\begin{figure}[tbp]  
  \begin{center}
    \includegraphics[height=7cm,width=9cm,keepaspectratio,clip]
    {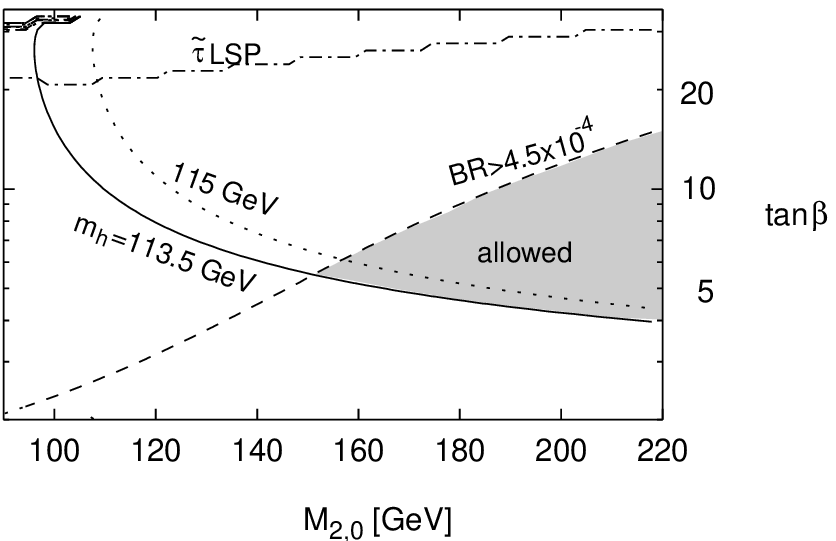}
  \end{center}
  \caption{The same as Fig.\ref{fig:Limit12R432R2posmu} but for $\mu < 0$.}
\label{fig:Limit12R432R2negmu}
\end{figure}

\begin{figure}[tbp]  
  \begin{center}
    \includegraphics[height=7cm,width=9cm,keepaspectratio,clip]
    {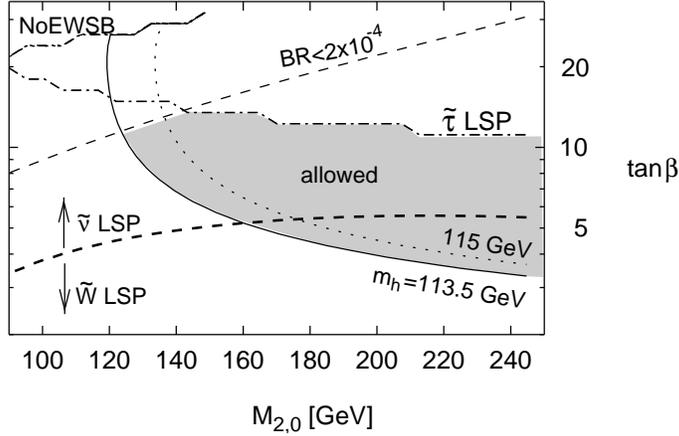}
  \end{center}
  \caption{Constraint in the $M_{2,0} - \tan \beta$ plane 
    for $M_{1,0}/M_{2,0}=2.5$, $M_{3,0}/M_{2,0}=1.5$, 
    $M_* = M_{\textrm{GUT}}$ and $\mu > 0$. 
    The regions above and below the thick dashed line, 
    the sneutrino and the Wino are the LSP, respectively.}
\label{fig:Limit12R2.532R1.5posmu}
\end{figure}

\begin{figure}[tbp]  
  \begin{center}
    \includegraphics[height=7cm,width=9cm,keepaspectratio,clip]
    {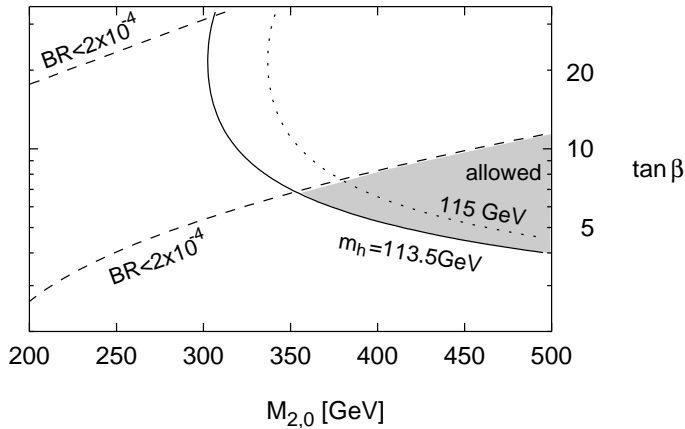}
  \end{center}
  \caption{Constraint in the $M_{2,0} - \tan \beta$ plane for the Higgsino 
    LSP case, $M_{1,0}/M_{2,0}=2$, $M_{3,0}/M_{2,0}=0.5$, 
    $M_* = M_{\textrm{GUT}}$ and $\mu > 0$. In the region between the two 
    dashed line, $\BRbsgamma$ is smaller than the lower limit of the CLEO result. }
\label{fig:Limit12R232R0.5posmu}
\end{figure}

\begin{figure}[tbp]  
  \begin{center}
    \includegraphics[height=7cm,width=9cm,keepaspectratio,clip]
    {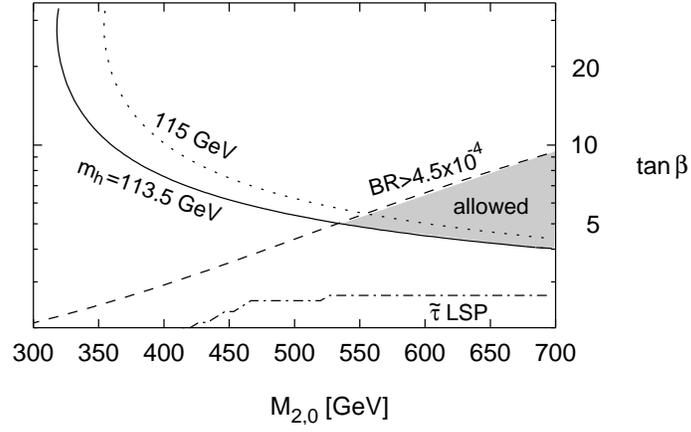}
  \end{center}
  \caption{The same as Fig.\ref{fig:Limit12R232R0.5posmu} but for $\mu < 0$.
    Notice that the region below the dash-dotted line is excluded by the 
    cosmological argument unlike in the other figures.}
\label{fig:Limit12R232R0.5negmu}
\end{figure}

The charged stau LSP can also be avoided if gaugino masses at the GUT scale 
are non-universal\cite{KomineYamaguchi}, i.e., 
the following boundary condition is given,
\begin{eqnarray}
  & & m_0^2=0, \quad A_0 = 0,  \nonumber \\
  & &M_1(M_{GUT}) = M_{1,0}, \quad M_2(M_{GUT}) = M_{2,0}, \quad 
  M_3(M_{GUT}) = M_{3,0} ~.
\end{eqnarray} 
This boundary condition can be given naturally within the GUT framework
\cite{Hisano-Goto-Murayama,non-universal-product-GUT}.
In this case, not only Bino-like neutralino, but also Wino-like, 
Higgsino-like neutralino or sneutrino can be the LSP.
For $M_{1,0}/M_{2,0} \gtrsim 2$ and  $M_{3,0}/M_{2,0} \gtrsim 1$, 
the LSP is wino-like neutralino.
For example, when $M_{1,0}/M_{2,0}=4$ and $M_{3,0}/M_{2,0}=2$, then
Wino mass and charged slepton mass are (notice that in this case the 
left-handed sleptons are lighter than right-handed sleptons);
\begin{eqnarray}
  M_2^2 \simeq 0.69 M_{2,0}^2, \quad 
  m_{\tilde{e}_L}^2 \simeq 1.06 M_{2,0}^2 -0.27 m_Z^2 \cos 2\beta.
\end{eqnarray}
The Higgsino is the LSP if $M_{3,0}/M_{2,0} \lesssim 0.5$.
For example, when $M_{1,0}/M_{2,0} = 2$ and $M_{3,0}/M_{2,0}=0.5$, then
the Higgsino mass and the right-handed slepton mass are
\begin{eqnarray}
  m_{\tilde{H}}^2 \simeq \mu^2 \simeq 0.416 M_{2,0}^2 -0.5m_Z^2, \quad
  m_{\tilde{e}_R}^2 \simeq 0.60 M_{2,0}^2 -0.23 m_Z^2 \cos 2\beta .
\end{eqnarray}
In the two cases given above, neutral wino or Higgsino is the LSP.
In fact from Fig.\ref{fig:Limit12R432R2posmu} - \ref{fig:Limit12R232R0.5negmu}
we find that neural particle is the LSP in large parameter region, 
thus it is cosmologically viable.

\section{Higgs mass and $\bsgamma$ constraint on No-scale scenario}
\label{sec:HiggsBsgamma}
In the previous section we take into account only LEP 2 bound 
and the cosmological constraint.
We find that the minimal scenario is severely constrained, 
but the other two scenarios are not.
In this section we also include the current Higgs mass bound and 
$\bsgamma$ constraint. 
As we will see, combining the above four constraint, 
not only the minimal case
but also the other two scenarios can be constrained more severely.
We also discuss the possibility whether this scenario can be seen 
at the Tevatron Run 2 or not.

Before we show the numerical results, some remarks on our calculation of the 
Higgs mass and $\BRbsgamma$ are in order.

It is well known that radiative correction is important when the lightest 
Higgs mass is evaluated \cite{OYY91}. 
In the present paper, 
the lightest Higgs mass is evaluated by means of the one-loop level 
effective potential\cite{1loopHiggs}.
This potential is evaluated at the renormalization point of the geometrical 
mean of the two stop mass eigenvalues 
$\sqrt{ m_{\tilde{t}_1} m_{\tilde{t}_2}}$.
We compared our result with a two-loop result 
by using {\it FeynHiggs}\cite{2loopHiggs},
and checked that the difference between these two 
results is smaller than 5 GeV as long as $\tan \beta$ is bigger than 5.
When $\tan \beta$ is close to 2, the difference can be 7 GeV.
However, as we will see later, Higgs mass bound plays an important rule around 
$\tan \beta \simeq 10$. 
And the two-loop effects always make the Higgs mass lighter than that obtained 
at the one-loop level.
So our conclusion is conservative and is not significantly changed 
by the two-loop effect.
We exclude the parameter region where the lightest Higgs mass is lighter than 
the current 95\% C.L. limit from LEP 2 experiments, 
$m_h > 113.5 \textrm{GeV}$ \cite{LEPHiggs}.

In the present paper, $\BRbsgamma$ is calculated including leading order (LO) 
QCD corrections\cite{Bertolini_etal}, 
and compare it to the current CLEO measurement. 
In the MSSM, chargino contribution can be comparable to the SM and 
charged Higgs contributions.
They interfere constructively (destructively) each other 
when $\mu<0$ ($\mu>0$).
The difference between the LO and the next-to-leading order (NLO) result can 
be sizable only when cancellation among different contributions at the LO is 
spoiled by the NLO contributions.
As we will see, however, the $\bsgamma$ constraint is severe when the 
interference is constructive.
In the case of destructive interference where the deviation from the NLO 
result may be large, this constraint is not so important.
Hence we expect that our conclusion is not changed significantly 
by the inclusion of the NLO corrections.
For the experimental value, we use 95\% C.L. limit from CLEO,  
$2.0 \times 10^{-4} < \BRbsgamma < 4.5 \times 10^{-4}$ \cite{CLEO99}.

\subsection{Minimal scenario}
First we show the numerical results for the minimal case. 
The case for $\mu > 0$ is shown in Fig.\ref{fig:Limitmimposmu}.
In this case, for small $\tan \beta$ region, 
the stop mass is not so large that 
radiative correction factor $\log( m_{\tilde{t}_1} m_{\tilde{t}_2}/ m_t^2)$
which raises the Higgs mass is small.
(For example, $m_{\tilde{t}_1}=361$ GeV and $m_{\tilde{t}_2}=567$ GeV 
for $M_{1/2}=200$ GeV and $\tan \beta = 3$).
Hence the Higgs mass limit constrains this scenario severely.
In Fig.\ref{fig:Limitmimposmu}, the Higgs mass bound and $\BRbsgamma$ 
constraints in the $M_{1/2} -\tan \beta$ plane are shown.
The regions below the solid line and above the dashed line are excluded by 
the Higgs mass and $\BRbsgamma$ bound, respectively.    
The indication of $m_h=115\GeV$ reported by LEP 2\cite{LEPHiggs} is also shown 
in this figure.
From the figure we find that the Higgs mass bound almost excludes the region 
where the stau LSP is avoided.
Note that, as we discussed earlier, the bound we put on the Higgs mass may be 
conservative, because the two loop correction may further reduce 
the Higgs mass.

The same figure but for $\mu < 0$ is shown in Fig.\ref{fig:Limitmimnegmu}.
Now $\BRbsgamma$ also constrains parameter region strongly since chargino 
contribution to $b \to s \gamma$ interferes with SM and charged Higgs ones 
constructively.
The region above the dashed line is excluded by $\BRbsgamma$ constraint.
We find that only one of the two constraints is enough to exclude all the 
region where cosmological bound and the smuon mass bound are avoided.
Hence if R-parity is conserved, i.e., the cosmological bound is relevant,
this scenario with $\mu < 0$ is excluded.

\subsection{$M_* > M_{\textrm{GUT}}$ case}
Next we show the numerical results in the case that the cutoff scale is 
larger than the GUT scale. 
As a typical example, 
we choose the minimal SU(5) as the theory above the GUT scale.
In Fig.\ref{fig:LimitMBC1e17posmu} and \ref{fig:LimitMBC1e17negmu}, 
results are shown for positive and negative $\mu$, 
respectively. In both figures, we take $M_* = 10^{17}$ GeV.
For $\mu > 0$ case, large parameter region is allowed and SUSY scale 
$M_{1/2}$ can be as small as about 180 GeV, which indicates the LSP mass 
$\tilde{\chi}^0_1 \simeq 80$ GeV.
For $\mu < 0$, as in the minimal case, 
$\BRbsgamma$ constraint is severer, and $M_{1/2}$ must be larger 
than around 280 GeV.
We also considered other values of the boundary scale $M_*$ from 
$5 \times 10^{16} \GeV$ to $10^{18} \GeV$, and checked that the behavior of 
the contour plot does not change so much. 

According to Ref.\cite{TevatronRun2}, Tevatron Run 2 experiment can explore 
up to $M_{1/2} \simeq 230$ GeV for integrated luminosity $2 \textrm{fb}^{-1}$.
Hence if $180 \GeV \lesssim M_{1/2} \lesssim 230 \GeV$ and $\mu >0$,
SUSY particles can be discovered at the experiment.
In this range, trilepton from chargino-neutralino associated production
$q \bar{q}^{\prime} \to \Neutwo \Chaone$, 
$\Neutwo \to \tilde{\ell}_R \ell \to \Neuone \ell \ell$, 
$\Chaone \to \Neuone \ell \nu$ is one of clean signals for SUSY search.
Notice that now two body decay 
$\tilde{\chi}^0_2 \to \tilde{\ell}_R \ell$ opens.
So same flavor, opposite sign dilepton from $\Neutwo$ decay may be useful.
The two body decay allows us to observe the peak edge of invariant mass 
of two leptons at the $M_{\ell \ell ~max}$. 
It is expressed in terms of the neutralino masses and the slepton mass as,
\begin{equation}
  M_{\ell \ell ~max} = m_{\tilde{\chi}^0_2} 
  \sqrt{1-\frac{m_{\tilde{\ell}_R}^2}{m_{\tilde{\chi}^0_2}^2}}
  \sqrt{1-\frac{m_{\tilde{\chi}^0_1}^2}{m_{\tilde{\ell}_R}^2}} ~.
\end{equation}
In table \ref{tab:Mllmax}, the dependence of $M_{\ell \ell ~max}$ on $M_*$ 
is shown. Here we fix $m_{\Neuone} = 100 \GeV$. 
Notice that as $M_*$ changes, the right-handed mass changes sizably
while the neutralino masses do not.
Hence we can obtain the mass relation among them and also cutoff scale $M_*$,
which corresponds to the compactification scale in the sequestered sector 
scenario, by measuring $M_{\ell \ell ~max}$. 
On the other hand, since only $M_{1/2} \gtrsim 280$ GeV is allowed for
 $\mu<0$, the Tevatron Run 2 
can not survey this scenario, and we have to wait LHC experiment.

\subsection{Case of non-universal gaugino masses}
Next, we turn to the case that gaugino masses are
non-universal at the GUT scale.
We explore the following three cases, 
Wino-like neutralino LSP, Higgsino-like neutralino LSP 
and the tau sneutrino LSP.
We will see that in the Wino-like neutralino LSP and tau sneutrino LSP cases,
constraint is not so severe even if we combine Higgs mass bound and 
$\BRbsgamma$ data,
but in the Higgsino-like LSP case where stops are as light as sleptons and 
charginos, the predicted Higgs mass tends to be small, 
and thus the Higgs mass bound becomes important.

First, we discuss the Wino-LSP case.
The results for $M_{1,0}/M_{2,0}=4$ , $M_{3,0}/M_{2,0}=2$ are shown in 
Fig.\ref{fig:Limit12R432R2posmu} and Fig.\ref{fig:Limit12R432R2negmu}, 
for $\mu>0$ and $\mu<0$, respectively.
In this case, we obtain a relatively large Higgs mass since $M_{3,0}$ is large
and so are the masses of stops. 
Hence, for $\mu>0$, $M_{2,0}$ can be as small as 100 GeV at 
$\tan \beta \simeq 10$, 
where the mass of the LSP $\tilde{\chi}^0_1$ is about 90 GeV.
For $\mu<0$, though $\BRbsgamma$ constraint is slightly severer than in the 
$\mu>0$ case, 
$M_{2,0} \simeq 160 \GeV$ is allowed, which corresponds to 
$m_{\tilde{\chi}^0_1} \simeq 142 \GeV$.
Hence the Wino-LSP with mass around 100 GeV is allowed.
Examples of the mass spectrum in this case are listed as point A ($\mu>0$) 
and point B ($\mu<0$) in Table \ref{tab:spectrum}. 
At the both points, $M_{1/2}$ is chosen to be near the smallest value 
such that all constraints are avoided.

In general, however, masses of $\Neuone$ and $\Chaone$ are highly degenerate
when Wino is the LSP.
In fact, from Table \ref{tab:spectrum}, we see that the mass difference is 
less than 1 GeV.
Therefore a lepton from $\Chaone \to \Neuone \ell \nu$ is very soft
and trilepton signal search is not useful
because acceptance cut usually requires the smallest transverse momentum of 
the three leptons $p_T(\ell_3)$ to be larger than 5 GeV\cite{TevatronRun2}.
Recently collider phenomenology in such cases are studied in Ref.
\cite{WLSPTev2}. 
It is shown that certain range of $m_{\Chaone}$ and 
$m_{\Chaone} - m_{\Neuone}$, SUSY signals which are different from those 
in the minimal case can be detected.
The high degeneracy requires to include radiative corrections to calculate
$m_{\Chaone} - m_{\Neuone}$ \cite{MizutaNgYamaguchi}, 
which is beyond of this work.
It deserves detail study to estimate the mass difference in the scenario. 

Since the constraint for the sneutrino LSP case 
in the $M_{2,0} - \tan \beta$ plane is similar to those in the Wino-LSP case,
we show the result for $\mu>0$ only in Fig.\ref{fig:Limit12R2.532R1.5posmu}.
In the figure, we take $M_{1,0}/M_{2,0} = 2.5$ and $M_{1,0}/M_{2,0} = 1.5$.
Notice that the decomposition of the LSP depends on $\tan \beta$ and 
the sneutrino is the LSP for $\tan \beta \gtrsim 5$.
An example of the mass spectrum is listed as the Point C in 
Table \ref{tab:spectrum}.
In this case, trilepton signal comes from 
$q \bar{q}^\prime \to \Neutwo \Chaone $, 
$\Chaone \to \tilde{\nu}_{\tau} \ell$, 
$\Neutwo \to \ell \bar{\ell} \nu_\tau \tilde{\nu}_{\tau}$.
Since $m_{\Chaone} - m_{\tilde{\nu}_\tau} = 6 \GeV$, 
$p_T$ of a lepton from $\Chaone$ decay is small and this signal may be hard 
to be detected. 
We may need unusual trigger to explore this scenario.  

Next, we turn to the Higgsino LSP case.
Higgsino LSP scenario is realized when $M_{3,0}$ is smaller than 
half of $M_{2,0}$, which indicates that colored particles are lighter than 
in the universal gaugino mass case. 
Hence the one-loop correction to the Higgs potential 
which enhances the Higgs mass is 
small and the Higgs mass constraint is important.  
The same figures as Fig.\ref{fig:Limitmimposmu} and  \ref{fig:Limitmimnegmu} 
are shown in Fig.\ref{fig:Limit12R232R0.5posmu} ($\mu>0$) and 
Fig.\ref{fig:Limit12R232R0.5negmu} ($\mu<0$)
for $M_{1,0}/M_{2,0}=2$ and $M_{3,0}/M_{2,0}=0.5$.
In order to satisfy the Higgs mass bound, 
$M_{2,0}$ must be larger than around 300 GeV.
Combining the bound with $\BRbsgamma$, constraint becomes severer, 
especially for $\mu<0$ case where $M_{2,0} \gtrsim 520 \GeV$ is required. 
Example of mass spectrum in this scenario is listed as Point D and E in  
Table \ref{tab:spectrum}. 
Again we choose almost the smallest value of $M_{2,0}$
where all constraints are avoided.
We see that the LSP mass must be at least $m_{\Neuone} \simeq 245 \GeV$ 
for $\mu>0$ and $m_{\Neuone} \simeq 370 \GeV$ for $\mu<0$.
Hence this scenario can not be explored at the Tevatron Run 2.

\section{conclusions}
\label{sec:conclusions}

The no-scale type boundary conditions are obtained 
in various types of SUSY models.
This scenario is attractive because it is highly predictive and can be a 
solution to the SUSY flavor problem.
In this paper we investigated the indication of the current Higgs mass and
$\bsgamma$ constraint on SUSY models with the boundary condition.

First we considered the minimal case where the universal gaugino mass are 
given at the GUT scale. 
This scenario has been already constrained by direct search at LEP and 
the cosmological bound severely, under the assumption of the exact R-parity. 
We showed that the Higgs mass bound and $\bsgamma$ constraint are also 
taken into account, then almost all the parameter region is excluded, 
leaving very narrow allowed region for $\mu > 0$.

Next we considered the case that the boundary condition is given above the GUT 
scale. 
Since the cosmological constraint is not severe, 
wide region of the parameter space is allowed.
In the $\mu>0$ case, Tevatron have a chance to observe SUSY signatures like 
trilepton events.
The scale $M_*$ may be explored by measuring the peak edge of invariant mass 
of two leptons at the $M_{\ell \ell ~max}$.
However for the $\mu<0$ case, since $M_{1/2} \gtrsim 280 \GeV$ is 
required, we have to wait LHC. 

Finally we considered the case where non-universal gaugino masses are given 
at the GUT scale.
We see that the Higgs mass bound is strong in the Higgsino LSP case
because stop masses are as light as sleptons and charginos.
The mass of the Higgsino-like neutralino must be larger than about 245 GeV 
and 370 GeV for $\mu>0$ and $\mu<0$, respectively.
In the Wino LSP and sneutrino LSP case, 
the mass of the LSP can be as small as 150 GeV.
However, the mass difference between the LSP and parent particles produced 
at the collider is much smaller than in the minimal case, 
unusual acceptance cut may be required.


\section*{acknowledgment}
The author would like to thank M. Yamaguchi for suggesting the subject, 
fruitful discussion and careful reading of the manuscript.
He also thanks T. Moroi and M. M. Nojiri for useful discussion.


\begin{table}
  \begin{center}
    \begin{tabular}{|r|ccc|}
      $M_*$ & $m_{\Neutwo}$ & $m_{\tilde{\ell}_R}$ & $M_{\ell \ell ~max}$
      \\ \hline
  $10^{17} \GeV $         & 185 GeV & 138 GeV & 85 GeV \\
  $2 \times 10^{17} \GeV$ & 185 GeV & 149 GeV & 81 GeV \\
  $4 \times 10^{17} \GeV$ & 186 GeV & 160 GeV & 74 GeV \\
  $10^{18} \GeV$          & 186 GeV & 172 GeV & 58 GeV \\
  $2.4\times10^{18}\GeV$  & 186 GeV & 182 GeV & 32 GeV \\
    \end{tabular}
  \end{center}
  \caption{The dependence of $M_{\ell \ell ~max}$ on $M_*$. 
    We fix $m_{\Neuone} = 100 \GeV$.}
  \label{tab:Mllmax}
\end{table}

\begin{table}
  \begin{center}
    \begin{tabular}{|c|ccccc|}
                       &Point A&Point B&Point C&Point D&Point E  \\ \hline
           $M_{1,0}$   & 480   & 680   & 375   & 800   & 1160    \\
           $M_{2,0}$   & 120   & 170   & 150   & 400   & 580     \\
           $M_{3,0}$   & 240   & 340   & 225   & 200   & 290     \\
         $\tan \beta$  & 10    & 6     & 10    & 7     & 6       \\
  $\textrm{sgn}(\mu)$  & $+$   & $-$   & $+$   & $+$   & $-$     \\ \hline

        $m_{\chi^0_1}$ & 91    & 142   & 112   & 225   & 371     \\
        $m_{\chi^0_2}$ & 199   & 290   & 156   & 277   & 402     \\
        $m_{\chi^0_3}$ & 366   & 520   & 343   & 338   & 491     \\   
        $m_{\chi^0_4}$ & 380   & 521   & 359   & 386   & 516     \\
        $m_{\chi^+_1}$ & 91    & 142   & 113   & 235   & 380     \\
        $m_{\chi^+_2}$ & 379   & 524   & 358   & 377   & 507     \\ \hline
    
     $m_{\tilde{e}_L}$ & 132   & 181   & 134   & 318   & 456      \\
     $m_{\tilde{e}_R}$ & 190   & 266   & 151   & 312   & 449      \\
  $m_{\tilde{\tau}_1}$ & 124   & 179   & 117   & 307   & 447      \\
  $m_{\tilde{\tau}_2}$ & 194   & 266   & 164   & 321   & 457      \\
     $m_{\tilde{\nu}}$ & 105   & 163   & 107   & 307   & 449      \\ \hline

     $m_{\tilde{u}_L}$ & 561   & 781   & 524   & 536   & 766      \\
     $m_{\tilde{u}_R}$ & 553   & 770   & 524   & 504   & 718      \\
     $m_{\tilde{d}_L}$ & 559   & 774   & 530   & 542   & 771      \\
     $m_{\tilde{d}_R}$ & 552   & 768   & 520   & 473   & 672      \\
     $m_{\tilde{t}_1}$ & 423   & 623   & 392   & 337   & 515      \\
     $m_{\tilde{t}_2}$ & 592   & 767   & 564   & 555   & 735      \\
     $m_{\tilde{b}_1}$ & 509   & 711   & 483   & 467   & 669      \\
     $m_{\tilde{b}_2}$ & 553   & 766   & 519   & 489   & 690      \\ \hline

               $m_h$   & 116   & 116   & 116   & 115   & 117

    \end{tabular}    
  \end{center}
  \caption{Examples of mass spectrum for five representative points.
    All dimensionful parameters are given in the GeV unit.}
  \label{tab:spectrum}
\end{table}

\end{document}